\documentstyle[prl,aps,multicol]{revtex}
\input{epsf}
\begin{document}
\draft
\twocolumn[\hsize\textwidth\columnwidth\hsize\csname @twocolumnfalse\endcsname
\title{Spin Polarized Ground State for Interacting 
Electrons in Two Dimensions}

\author{ Giuliano Benenti$^{(a,b,c)}$, Ga\"etan Caldara$^{(a)}$,
and Dima L. Shepelyansky$^{(a)}$}

\address{$^{(a)}$Laboratoire de Physique Quantique, UMR 5626 du CNRS, 
Universit\'e Paul Sabatier, 31062 Toulouse Cedex 4, France}
\address{$^{(b)}$International Center for the Study of Dynamical 
Systems, \\ 
Universit\`a degli Studi dell'Insubria, via Valleggio 11, 22100 Como, Italy}  
\address{$^{(c)}$Istituto Nazionale di Fisica della Materia, 
Unit\`a di Milano, via Celoria 16, 20133 Milano, Italy} 

\date{\today}

\maketitle

\begin{abstract}
We study numerically the ground state magnetization for clusters
of interacting electrons in two dimensions 
in the regime where the single particle wavefunctions
are localized by disorder. It is found that the Coulomb
interaction leads to a spontaneous 
ground state magnetization. For a constant electronic density, 
the total spin increases linearly with the number of particles, 
suggesting a ferromagnetic ground state in the thermodynamic limit.
The magnetization is suppressed when the single particle states
become delocalized. 

\end{abstract}
\vskip1pc]

\narrowtext

Ferromagnetic instabilities result from the interplay 
between the electronic Coulomb interaction and the Pauli 
principle.
In the Pauli picture, electrons populate the non-interacting orbitals
of a system, such as a quantum dot or a metallic 
grain, in a sequence of spin up - spin down electrons. 
The resulting minimum spin state  minimizes the kinetic energy:  
it costs energy to flip a spin since it must be promoted to 
a higher energy level. Thus the
total spin of the system is
$S=0$ when the number of electrons $N$ is even and 
$S=1/2$ at odd $N$.  In contrast, the maximum spin allows 
a maximally antisymmetric coordinate wavefunction, thus 
reducing the effect of the Coulomb repulsion (a familiar 
example of this is Hund's rule for atoms). 
This leads to the Stoner instability \cite{auerbach}, which 
gives a spontaneous magnetization when the typical 
interaction exchange energy between two particles 
close to the Fermi level is of the order of the single 
particle level spacing. 

Spontaneous ground state magnetization gives rise to 
interesting effects, which are in the focus of many 
recent studies. In quantum dots, a ground
state spin polarization can explain the absence of an
even-odd asymmetry in the addition spectra in the
Coulomb blockade regime \cite{berkovits,marcus,simmel}.
Also, the addition of an electron to the dot may flip
the spin of other electrons already in the dot: if the
total spins of the ground states of successive number of electrons
differ by more than $1/2$, spin selection rules
suppress the corresponding conductance peak (spin blockade)
\cite{dietmar}. Spontaneous magnetization effects could also 
explain the presence of kinks in the magnetic field dependence
of the Coulomb blockade peak positions \cite{baranger,altshuler}. 
The stability of the minimum spin ground state in a quantum
dot was analyzed for weak interactions in \cite{prus}. 
Within perturbation theory, the effective interaction strength 
is enhanced by the presence of disorder, leading to a
ferromagnetic instability already below the Stoner
threshold \cite{kamenev}. In the diffusive regime,
recent studies have also considered the effect of mesoscopic
wave function fluctuations \cite{brower} and of off-diagonal
interaction matrix elements beyond the mean field treatment
\cite{jacquod}. The appearance of local magnetic moments
has been discussed also in the strongly correlated limit,
at the quantum melting of the Wigner crystal \cite{franck}. 

Although the Stoner instability signals the presence of
short range magnetic ordering, it is not clear if it will
also lead to a ferromagnetic ground state in the thermodynamic
limit. Actually, the Stoner criterion is obtained within the
mean field Hartree-Fock approximation and overestimates the
long-range magnetic ordering, predicted also in one and two
dimensions for the Hubbard model at finite temperatures,
thus violating the Mermin and Wagner theorem \cite{auerbach}.
The possibility of a ferromagnetic phase in strongly
correlated two-dimensional (2D) systems was considered
in \cite{kelly,rajagopal} (see also Refs. \cite{senatore,ceperley}) 
and has recently received experimental
support in dilute 2D electron gases \cite{sarachik}.
At the same time, recent studies of fermionic models with 
random two-body interactions
show that the ground state polarization is strongly reduced 
by off-diagonal interaction matrix elements \cite{jacquod,kaplan}.

In this Letter, we investigate numerically the possibility
of a ferromagnetic ground state in the regime where
the single particle localization length is smaller than
the system size. Without interaction this condition is always
satisfied in two dimensions in the limit of large system size
\cite{abrahams}.

We study a disordered square lattice with $N$ fermions 
on $L^2$ sites. The Hamiltonian is defined by  
\begin{eqnarray} 
\label{hamiltonian} 
\hat{H}=-V\sum_{<{\bf i},\,{\bf j}>,\sigma} 
c_{{\bf i}\sigma}^\dagger c_{{\bf j}\sigma} 
+\sum_{\bf i\sigma} \epsilon_{\bf i} 
n_{{\bf i}\sigma} +
\nonumber 
\\
U_H \sum_{\bf i} n_{{\bf i} \uparrow} n_{{\bf i} \downarrow} 
+U \sum_{{\bf i} \neq {\bf j},\sigma,\sigma'} 
\frac{n_{{\bf i} \sigma} n_{{\bf j} \sigma'}}{|{\bf i}-{\bf j}|}, 
\end{eqnarray} 
where $c_{{\bf i}\sigma}^\dagger$ ($c_{{\bf i}\sigma}$) 
creates (destroys) an electron at site ${\bf i}$ with spin $\sigma$, 
$n_{{\bf i}\sigma}=c_{{\bf i}\sigma}^{\dagger}
c_{{\bf i}\sigma}$ is the corresponding 
occupation number,     
the hopping term $V$ between nearest neighbors characterizes 
the kinetic energy, 
random site energies $\epsilon_{\bf i}$ are taken from a box 
distribution over $[-W/2,W/2]$, 
$U_H$ and $U$ measure the strength of the on-site Hubbard 
interaction and of the Coulomb interaction, respectively, 
and $|{\bf i}-{\bf j}|$ is the inter-particle shortest distance 
computed on a 2D torus (periodic boundary conditions are 
taken in both directions). In the following we choose $U_H=U$.

The Hamiltonian (\ref{hamiltonian}) commutes with the total spin
($[\hat{S}^2,\hat{H}]=0$) and its component along an arbitrary
$z$-direction ($[\hat{S}_z,\hat{H}]=0$). 
Therefore $\hat{H}$ can be written in a
block-diagonal form, with $N+1$ blocks where
$S_z=-N/2,-N/2+1,...,N/2$, respectively. We consider the block
with $S_z=0$ only, since it is sufficient for analysis of the ground state  
magnetization. Indeed, due to spin rotational symmetry, the system
has a $2S+1$ degeneracy (where $S$ is the total spin), with 
$S_z=-S,...,+S$; therefore all the eigenvalues of the Hamiltonian
(\ref{hamiltonian}) belong to the spectrum of the $S_z=0$ subspace.
 
The numerical studies of  the model (\ref{hamiltonian}) at a finite 
density of interacting particles above a frozen Fermi sea are performed 
in the following way:
\newline\noindent 
(i) Single particle eigenvalues $\epsilon_\alpha$ and eigenstates 
(orbitals) $\phi_\alpha({\bf i})$ ($\alpha=1,...,L^2$) are 
obtained via numerical diagonalization of the 
Hamiltonian (\ref{hamiltonian}) at $U_H=U=0$.      
\newline\noindent 
(ii) The Hamiltonian (\ref{hamiltonian}) is written in 
the basis of non-interacting orbitals obtained in (i): 
\begin{eqnarray} 
\label{hamobb} 
\hat{H}=\sum_{\alpha,\sigma} 
E_\alpha d_{\alpha\sigma}^\dagger d_{\alpha\sigma}+ 
U_H\sum_{\alpha,\beta,\gamma,\delta} H_{\alpha\beta}^{\gamma\delta}
d_{\alpha\uparrow}^\dagger d_{\beta\downarrow}^\dagger  
d_{\delta\downarrow} d_{\gamma\uparrow}+
\nonumber
\\
U\sum_{\alpha,\beta,\gamma,\delta,\sigma,\sigma'}
C_{\alpha\beta}^{\gamma\delta}
d_{\alpha\sigma}^\dagger d_{\beta\sigma'}^\dagger  
d_{\delta\sigma'} d_{\gamma\sigma},
\end{eqnarray} 
with $d_{\alpha\sigma}^\dagger=\sum_{\bf i}\phi_\alpha({\bf i}) 
c_{{\bf i}\sigma}^\dagger$, and 
transition matrix elements  
\begin{equation}   
H_{\alpha\beta}^{\gamma\delta}= 
\sum_{\bf i} \phi_\alpha({\bf i})\phi_\beta({\bf i})
\phi_\gamma({\bf i})\phi_\delta({\bf i}),     
\end{equation}
\begin{equation}   
C_{\alpha\beta}^{\gamma\delta}= 
\sum_{\bf i\neq j} \frac{\phi_\alpha({\bf i})\phi_\beta({\bf j})
\phi_\gamma({\bf i})\phi_\delta({\bf j})}{|{\bf i}-{\bf j}|}.     
\end{equation}
\newline\noindent 
(iii) The Fermi sea is introduced by restricting the sums in 
(\ref{hamobb}) 
to orbitals with energies above the Fermi energy $\epsilon_{M_F}$: 
$\alpha,\beta,\gamma,\delta> M_F$. We consider a filling factor 
$\nu_F=M_F/L^2=1/4$ (corresponding to $2 M_F$ frozen electrons due
to spin degeneracy) and a finite density $\rho=N/L^2$ of $N$ interacting
particles above the Fermi level. The frozen Fermi sea approximation
is introduced for the sake of simplicity, since it allows us to
avoid the band tail, where the single particle density of states
and the one body localization length have a strong energy dependence.
The advantages of such an approach have been demonstrated in
\cite{eplpichard}. However,
we have also checked that the results presented in this Letter
are qualitatively similar when $M_F=0$.
\newline\noindent 
(iv) The basis of the Slater determinants, built from the single particle 
orbitals $\phi_\alpha$, is energetically truncated at high 
energy orbitals by means of the   
condition $\sum_{i=1}^{N} (m_i-M_F)\leq M$. Here $m_i$ is
the orbital index 
for the $i$-th quasiparticle ($m_i>M_F$). The truncated Hamiltonian
still commutes with the total spin when $S_z=0$.
\newline\noindent
(v) We diagonalize the many-body truncated Hamiltonian.
Then the total spin of a given eigenstates $|\Psi_i\rangle$
is found via the application of the operator
$\hat{S}^2$: $\hat{S}^2|\Psi_i\rangle=S_i(S_i+1)|\Psi_i\rangle$
(we take $\hbar=1$). 

We consider $N=2,4,6,8$ particles on a square lattice of
size $L=8,11,14,16$ respectively, at an approximately constant density 
$\rho=N/L^2\approx 1/32$,  $0.5\leq U/V \leq 2$, 
$2\leq W/V \leq 10$.
In particular, we focus on the localized regime $W=10V$,
where the single particle localization length
$l_1\approx 4 < L$. Data are averaged over a number of 
disorder configurations between $200$ and $5000$.  

\begin{figure}
\epsfxsize=8cm
\epsffile{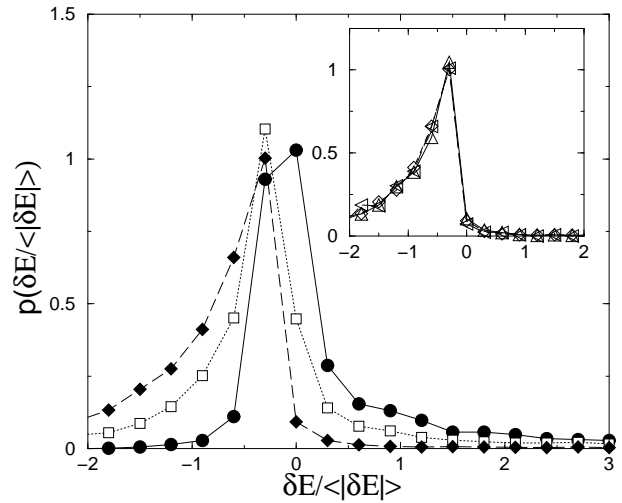}
\caption{ 
Normalized distribution of the energy differences $\delta E$ between the 
ground state energies $E_0(S)$ in the spin sectors
$S=1$ and $S=0$. Here $\delta E = E_0 (S=1) - E_0 (S=0)$
and $<|\delta E|>$ is the absolute value of $\delta E$
averaged over disorder configurations. Data are shown
for $N=8$ particles 
on a square lattice of size $L=16$, disorder strength $W=10V$, 
and interaction strengths $U=0.5 V$ (circles), $U=V$ (squares), 
and $U=2 V$ (diamonds). 
Inset shows the  distribution at $U=2V$ as a function of the size 
of the truncated  Hilbert space: $N_H=932$ (triangles up), 
$2097$ (diamonds), and $4354$ (triangles left).}    
\label{fig1}
\end{figure}

The distribution of the energy differences $\delta E$ between 
the ground state energies $E_0(S)$ in spin sectors 
$S=1$ and $S=0$ is shown in Fig. \ref{fig1} for $N=8$ particles in 
the localized regime with $W=10V$ ($\delta E = E_0 (S=1) - E_0 (S=0)$). 
One can see the mesoscopic Stoner mechanism: 
electron-electron interactions give a 
spontaneous magnetization ($\delta E<0$), 
with a probability to have a polarized state 
increasing with the growth of interaction.
The inset of Fig. \ref{fig1} demonstrates that 
our results are stable when the size $N_H$ of the 
truncated Hilbert space is changed by a factor of five. 
Even though we cannot exclude the existence of very 
slow $N_H$-variations this check shows that the
truncation does not significantly affect the
ground state polarization.

\begin{figure}
\epsfxsize=8cm
\epsffile{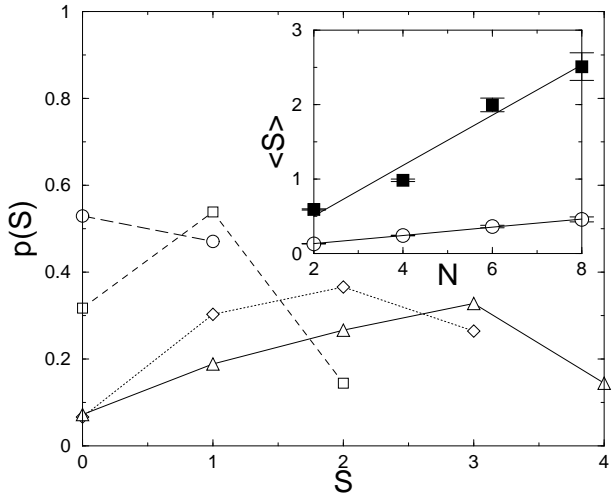}
\caption{Probability distribution $p(S)$ of the ground state spin 
for $W=10V$, $U=2V$. Here, $N=2$, $L=8$ (circles), 
$N=4$, $L=11$ (squares), $N=6$, $L=14$ (diamonds), 
$N=8$, $L=16$ (triangles). 
Inset: ground state average spin as a function of the number 
of particles for $U=0.5 V$ (circles) and $U=2 V$ (squares). 
Straight lines give linear fits, with $<S (U=0.5)>\approx 0.06 N$
and $<S(U=2)>\approx 0.34 N$. Here and in the next figures the error-bars 
show the size of statistical errors.
}      
\label{fig2}
\end{figure}

The main result of our Letter is shown in Fig. \ref{fig2}:
when the number of particle is increased (at a constant 
electronic density) the ground state can be found with 
high probability at larger and larger spin values. 
In the inset one can see that the ground state average 
magnetization increases linearly with the number of particles, 
$<S>\approx \alpha(U) N$ \cite{noteg}, with the slope $\alpha(U)$ 
growing with $U$, which determines the strength of the 
interaction exchange term. 
The extrapolation of the results presented in this figure
would give a ferromagnetic ground state in the 
thermodynamic limit. 

\begin{figure}
\epsfxsize=8cm
\epsffile{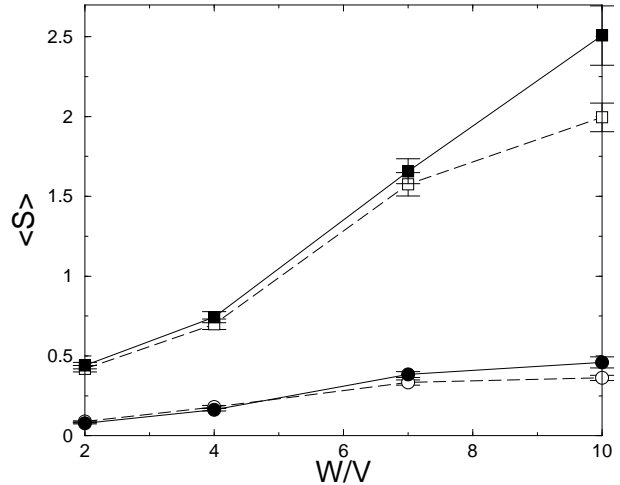}
\caption{Ground state average spin $<S>$ as a function of disorder
strength $W/V$, 
for $U=0.5 V$ (circles) and $U=2 V$ (squares); $N=6$ 
(empty symbols) and $N=8$ (filled symbols).}  
\label{fig3}
\end{figure}

The dependence of the average magnetization on the 
disorder strength is shown in Fig.\ref{fig3}. 
One can see that for sufficiently  
strong interaction disorder favors ground state
spin polarization. Indeed, for $U/V=0.5$
at strong disorder, $W/V=10$,
the total spin remains less than $0.5$
while for $U/V=2$ it becomes five time larger.
The significant average magnetization $<S>$
appears in  the 
localized single particle phase ($W/V=7,10$), 
while in the delocalized 
regime ($W/V=2,4$) it remains rather weak.

\begin{figure}
\epsfxsize=8cm
\epsffile{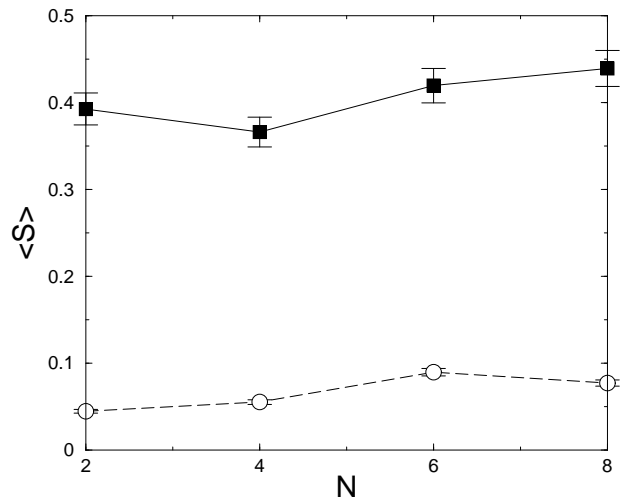}
\caption{Ground state average spin $<S>$ as a function of the 
number of particles, for $W= 2V$, $U=0.5V$ (circles) and 
$U=2V$ (squares). The system size
changes from $L=8$ for $N=2$ to $L=16$ for $N=8$.}  
\label{fig4}
\end{figure}

This is further confirmed in Fig. \ref{fig4}, which shows 
the size dependence of the average spin in the delocalized 
regime $W=2V$.  With the change of  the number of particles  
between $N=2$ and $N=8$ the average
magnetization  $<S>$ remains constant, 
with non monotonous fluctuations around its average value.
This is in a sharp contrast with magnetization behavior
in the localized phase (see inset in Fig.~2)
where $<S>$ demonstrates a monotonous growth with $N$. 

The ensemble of our results allows us to propose the following 
physical scenario. In the regime when the
Fermi energy is larger than the electron-electron interaction
the perturbation theory \cite{kamenev}
tells that the correction to the spin susceptibility $\delta \chi$
induced  by interaction is inversely proportional to the
conductance of the sample $g$: $\delta \chi \propto 1/g$.
With the increase of disorder $g$ drops and becomes of the order of
one when the single particle localization length $l_1$
becomes comparable with the sample size. This indicates
that the spin effects become more important in the 
regime of strong disorder, that is in agreement with our results
(see Fig.~3).  
In the nonperturbative diffusive regime, non-diagonal 
interaction matrix elements start to give quantum fluctuations 
beyond mean-field Stoner approach. These interaction fluctuations 
favor small spin values, since the number of off-diagonal 
scattering events is larger in the lower spin sectors of 
the Hilbert space. This effect can prevent the ground state spin 
from achieving a full polarization \cite{jacquod}. 
On the contrary, in the localized regime the 
off-diagonal fluctuations are strongly reduced: due to single 
particle exponential localization, Coulomb repulsion can 
induce electron jumps only inside the localization domains, 
all the other scattering events giving an exponentially 
small contribution. 
Therefore, a possible scenario is the following: 
Stoner instability gives, at strong enough interaction 
and/or disorder, spin polarization in domains of the size of the 
single particle localization length, then the coupling between these 
domains gives global magnetization. 
The long range nature of the Coulomb interaction
seems to play a crucial role in this physical picture.
Indeed, recent quantum Monte Carlo studies
of the ground state magnetization in the Hubbard model
with disorder show disappearance of any magnetic order
at strong disorder \cite{hub}.

In summary, we have shown that Coulomb repulsion can lead to 
spontaneous ground state magnetization. In the regime with localized
single particle wavefunctions, the total spin increases linearly with 
the number of particles. Even though we cannot exclude that this magnetic 
ordering can become limited at some large finite sizes, 
our results suggest the appearance 
of a ferromagnetic ground state 
induced by disorder and localization in the thermodynamic limit. 

We thank the IDRIS in Orsay and the CalMiP in Toulouse for access to 
their supercomputers.


\begin{thebibliography}{99}
\bibitem{noteg} The average spin $<S>$ is defined by the relation
$ \langle\langle\Psi_i|  \hat{S}^2|\Psi_i\rangle\rangle =<S>(<S>+1)$,
where $ \langle\langle ...\rangle\rangle$  
marks the averaging over the ground state and disorder configurations.
\end{thebibliography}
\end{document}